\documentclass[10pt,conference]{IEEEtran}
\usepackage[table]{xcolor}
\usepackage{booktabs,array,xcolor,colortbl,multicol,multirow,epsfig,graphicx,subfigure,amsmath,amsthm,amsfonts,amssymb,bm,algorithmic,algorithm}
\usepackage{caption}
\ifCLASSINFOpdf
\else
\fi
\begin{document}
%
\title{Sales pipeline win propensity prediction: a regression approach}

\author{\IEEEauthorblockN{Junchi Yan}
\IEEEauthorblockA{
IBM Research - China\\
Shanghai, China 201203\\
yanjc@cn.ibm.com}
\and
\IEEEauthorblockN{Min Gong}
\IEEEauthorblockA{IBM Research - China\\
Shanghai, China 201203\\
gminsh@cn.ibm.com}
\and
\IEEEauthorblockN{Changhua Sun}
\IEEEauthorblockA{IBM Research - China\\
Beijing, China 100193\\
schangh@cn.ibm.com}
\and
\IEEEauthorblockN{Jin Huang}
\IEEEauthorblockA{IBM Research - China\\
Shanghai, China 201203\\
huangjsh@cn.ibm.com}
\and
\IEEEauthorblockN{Stephen M. Chu}
\IEEEauthorblockA{IBM Research - China\\
Shanghai, China 201203\\
schu@us.ibm.com}

}


%


\maketitle

\begin{abstract}
Sales pipeline analysis is fundamental to proactive management of an enterprize's sales pipeline and critical for business success. In particular, win propensity prediction, which involves quantitatively estimating the likelihood that on-going sales opportunities will be won within a specified time window, is a fundamental building block for sales management and lays the foundation for many applications such as resource optimization and sales gap analysis. With the proliferation of big data, the use of data-driven predictive models as a means to drive better sales performance is increasingly widespread, both in business-to-client (B2C) and business-to-business (B2B) markets. However, the relatively small number of B2B transactions (compared with the volume of B2C transactions), noisy data, and the fast-changing market environment pose challenges to effective predictive modeling. This paper proposes a machine learning-based unified framework for sales opportunity win propensity prediction, aimed at addressing these challenges. We demonstrate the efficacy of our proposed system using data from a top-500 enterprize in the business-to-business market.
\end{abstract}

%
\IEEEpeerreviewmaketitle
\section{Introduction}
A sales strategy involves defining a sales process that accurately reflects a company's customers and the products or solutions that it sells. By aiming to truly understand its customers and address their problems, a company can define and execute a sales process that increases the likelihood of achieving the company's financial objective. Traditionally, sales departments have operated with no formally defined sales strategy or sales process, In many cases, each salesperson developed an individual, non-documented, and personalized selling approach with few to no metrics in place to measure the performance or success objectively and quantitatively \cite{vavricka1997implementing}.

With the development of information technology, companies are adopting more systematic and digitalized sales management systems\cite{selden1996sales,selden2000power} to support the sales process. A common approach, known as the opportunity management process, is as follows. As new sales opportunities (``lead'') are identified, the salesperson (``seller'') enters these leads into the sales opportunity pipeline management system. This seller is often referred to as the ``opportunity owner''. The opportunity pipeline management system provides the company with a holistic view of the status of each opportunity in the sales pipeline\cite{selden1996sales}. The company uses sales stages and other attributes associated with each lead to record, manage and track leads in the sales pipeline. A lead's sales stage characterizes how far the lead has progressed through the sales process. A key objective of the opportunity management process is to optimize the yield of the lead pool, i.e., to increase the number of leads that are converted to sales. Companies typically perform extensive and continual analysis of the leads in the sales pipeline to assess the number of leads at each sales stage, the efficiency of each seller and sales team, and the likelihood that leads will be converted to sales. Companies also analyze the pipeline to rate performance at different levels of the organizational hierarchy such as different geographies or different product lines. In particular, before it arrives at the won status, in other stages, it still runs the risk of losing the deal.

Digitizing sales pipeline information provides a first step toward improved pipeline management, more reliable yield prediction, and better evaluation of the quality of the pipeline. However, questions arise as to how the massive volumes of data collected in the sales opportunity pipeline management system can be effectively used to support sellers in the sales process to improve the yield rate. For example, under the tremendous pressure to produce revenues sellers often resort to working any leads at-hand, irrespective of the quality of these leads. Further, it is difficult for sellers to estimate lead quality based upon lead attributes in the pipeline database especially for those who are newly recruited freshmen sellers.

Moreover, the synergy between sales field team and marketing needs to be effectively integrated to ensure the campaign can lead to actual sales. Marketing is the pre-pipeline work to find and develop prospects before they are entered into the pipeline as live opportunities. Many companies dedicate entire departments and numerous processes to marketing the business\cite{coe2004fundamentals} in an attempt to generate and qualify leads. However, sales and marketing teams may be unable to efficiently and effectively manage marketing activity and subsequent leads. Due to a lack of evaluation and modeling how the marketing factors will influence the win propensity, there is often a risk of significant waste of marketing dollars because many sales leads are never worked by the sales force. This poses the request to better combine and analyze marketing information with the sales leads data.

The above observations highlight the role played by win propensity estimation as direct support in the sales process. More specifically, the sales lead prediction problem can be formally defined as follows. For any on-going sales lead in the sales pipeline, quantitatively estimate the probability that the lead will be won prior to the end of the current business cycle (e.g., prior to the end of the current quarter). Currently, using business knowledge and experience, a field sales manager may expend significant effort manually classifying the quality of all leads in his team's sales pipeline. This classification is then used to assign sales effort. However this method is unsystematic, and personal bias is unavoidable, making comparison of lead quality difficult when comparing lead ratings from multiple sales teams (since some teams may overestimate and some teams may underestimate lead quality). More importantly, sales forecasting is highly complex due to the influence of internal and external factors that may contribute to a lead's win propensity in a nonlinear manner\cite{kuo2001sales}. These complex factors are difficult to capture by documented knowledge, and may also evolve over time due to changes in product lines, newly acquired brands, or the emergence of new markets due to emerging sales branch planning and expansion.

Over the past three decades, few statistical and machine learning methods have been adopted to address the enterprize sales forecasting problem: \cite{armstrong1972comparative,yokuma1995beyond,davis2007organizational,lee2012hybrid}. Specifically, \cite{armstrong1972comparative,yokuma1995beyond} evaluate different methods for sales forecasting; in both of these papers forecasting is performed at an aggregate level rather than a lead level and the methods are broadly classified into subjective methods and objective methods or, from another perspective, naive methods and causal methods. For instance, \cite{armstrong1972comparative} studies forecast accuracy, and \cite{yokuma1995beyond} studies another criteria. \cite{davis2007organizational} reports on the results of empirical case studies performed among eighteen global manufacturing firms and advocates organizational factors that are key to sales forecasting. In \cite{lee2012hybrid}, a hybrid learning-based forecasting approach is proposed by combining multiple predictive models with many other models.

Our approach to sales lead prediction involves two steps. The first step involves collecting the training dataset including historical leads, associated profile features of these leads, and labels which are win or non-win outcome in the end of an time window. In practice, the time window usually refers to a fiscal quarter as a business cycle. This dataset is used to build the predictive model as follows. All leads in the pipeline at time point $t_s$ are identified and included in the dataset together with the features of these leads. This dataset is used as input to train the model. At time $t_e>t_s$, these leads are again reviewed to generate their labels, based on the state of the lead at time $t_e$. Common classifications are 'win', 'lose', or 'pending' (i.e., the company has won the lead, the company has lost the lead, or final disposition of the lead is not yet known.) The predictive model is trained to learn the mapping between the input features and the categorical labels. Since there is high focus on leads that are won, the problem can be further simplified to a binary classification task -- to identify leads that are won from leads in non-won states (lost or pending).

The second step involves using the trained model to estimate the win propensity for leads that are in a pending status, usually for current time point. The user also specifies a time frame of interest within which the lead should be won. The win propensity is typically aligned with business planning cycles. For most listed companies, sales and financial performance are evaluated on a quarterly basis. As such, most companies are interested in predicting win propensities for given quarters.

In the third step, two metrics are used to evaluate the performance of the predictive model, including ROC score and Gain score. This evaluation can be performed when the actual outcome comes up in the end of the time window.
\section{Features extraction and modeling}
A variety of information can be explored as input features to build a computational model under supervised learning/semi-supervised methodology \cite{TongICIG11,YanVCIP11,YanEL11,YanICIP10,YanSPL10} meanwhile enhancing its robustness against noises \cite{LiECCV10,LiICIP10}. We propose a systematic method to design the feature extraction mechanism for sales lead analytics. The derived features can be categorized to unary and interaction ones.
\begin{table}[t]
\center
\small
\caption{Exemplary profile features associated with the sales leads. Different types of features are considered.}
\begin{tabular}{|l|l|l|}
    \hline
    Profile&type&remark or examples\\\hline
    geography & categorical & Greater China, Southeast Asia\\\hline
    deal size	&categorical &expected deal size in USD\\\hline
sector&	categorical &general business, industry clients\\\hline
new client&	flag &first-time purchase or not\\\hline
lead age	&continuous &days since the opp. being created\\\hline
industry&	categorical &health-care, energy and utility\\\hline
product&	categorical&Sub-brands of the main brand\\\hline
\end{tabular}\label{tab:features}
\end{table}

\textbf{Unary features} Unary features are the raw features associated with the leads such as client type, sector, lead age (created date to now), and sales stage. Unary features are sales lead attributes that are typically found in the sales lead table in the sales pipeline data warehouse.  Our propensity modeling approach makes use of lead-profile, product-profile, seller-profile, and customer-profile data. In the Applications and Experiments section below, we describe the contribution of these features to the win propensity prediction. Readers are referred to Table.\ref{tab:features} for the description of exemplary features.

\textbf{Interaction features} One major technical problem is the inherent difficulty inside the quarterly prediction problem. Strictly speaking, there is an additional variable associated with the outcome, i.e. time stamp. While in our binary supervised learning paradigm, we ignore this fine-grained time resolution because it is generally assumed the sales resource and sellers' actions are driven by the quarterly set target thus it becomes almost unpredictable and mathematically difficult to infer the outcome distribution in the whole quarter. On the other hand, intuitively, there does exist the pattern that as it approaches the quarter end, the chance to win the lead becomes smaller and smaller. Another practical consideration comes from the way of building the training set. Every week we obtain the updated copy of the lead data. For a specified lead, it would experience $k$ weeks (from week $i$ to week $i+k-1$) before it arrives its outcome in week $i+k$. Accordingly, a series of training data copies are generated whose input features are recorded as $f_i, f_{i+1}, \ldots, f_{i+k-1}$. One shall note that these input features keep unchanged in most cases thus only reproduce replicates of training samples if we ignore the time stamp associated with the features. To better model the time-related information, we propose to derive additional interaction features by multiplying the week number with other static profiles including sales stage, opportunity owner, current quarter, lead age etc. We found this feature design mechanism has two advantages: i) for model training: all snapshots for the same lead can be used appropriately, instead of generating redundant replicated training samples; ii) for model prediction: since the model is trained using all copies of the same lead by considering different time stamps, the scoring would be stable for a given lead as week continues. Otherwise, if we train each model for each week separately, this would cause possible scoring fluctuation because the models used for scoring are different from current week to next one.

\textbf{Modeling} Under the supervised learning based classification paradigm, there is a broad set of models can be used. For instance, the Support Vector Machine (SVM), the Artificial Neural Network (ANN), and the Logistic Regression (LR) and so forth. Another thread of research is to ensemble the predictive power of different models into a unified and boosted model, based on the assumption that the input models are heterogenous and have different discriminances for different testing samples. In our solution, we chose to use the Logistic Regression (LR) model for two main reasons: i) LR can directly produce probability instead of a score which lacks of direct business interpretation; ii) LR is very efficient as we compare with other models, especially a recent four years of worldwide data is used for model training. Furthermore, as we encode additional interaction variables, which make the size of input feature set significantly larger than the raw unary feature set. As a result, the training process becomes more time-costive which calls for efficient training algorithm; iii) last but not least, the business side needs a user-friendly model and LR is very intuitive due to its linear nature.

\section{Deployment and experiments}
We perform our study on a Fortune 500 multinational technology company in the B2B market environment. Throughout this section, due to the sensitivity of the proprietary company-owned selling data, we de-identified the brand name and other profile information, only leave relative metrics such as AUC score. More specifically, we confine our study on the company's service lead data instead of the whole product lines in this case study. This dataset covers various geometry-wise unit including Greater China, Japan, India etc.

Receiver operating characteristic (ROC) is widely used to measure the performance of a predictive model, as it is insensitive to imbalanced two-class sample distribution and free from cut-off point selection for sample labeling as done in recall and precision calculation \cite{brown2006receiver,fawcett2006introduction}. While in real business case, the sales decision makers are more interested in the cumulative gain curve \cite{jarvelin2002cumulated} and its associated gain score which examines the actual win case distribution on the prediction output list ranked by propensity score. In the sales prediction context, the x-axis shows the percentage of opportunities contacted, which is a fraction of total cases based on the win propensity score ranking. In its original definition, y-axis shows the percentage of covered total actual win cases by the contacted customers so far in x-axis. This is a percentage of the total possible positive responses i.e. the number actual win cases. Thus we use the gain score as the performance metric for its interpretation power.

Before diving into evaluation, we first present several application tools derived by the quarterly predictive model.

\textbf{Web portal} This portal shows the pipeline quality and gap analysis against the specified quarterly quota target. The heat map that covers various areas and service lines, by computing the expected revenue, is mainly used by sales leaders.

\begin{figure}[t]
\centering
\includegraphics[width=0.15\textwidth]{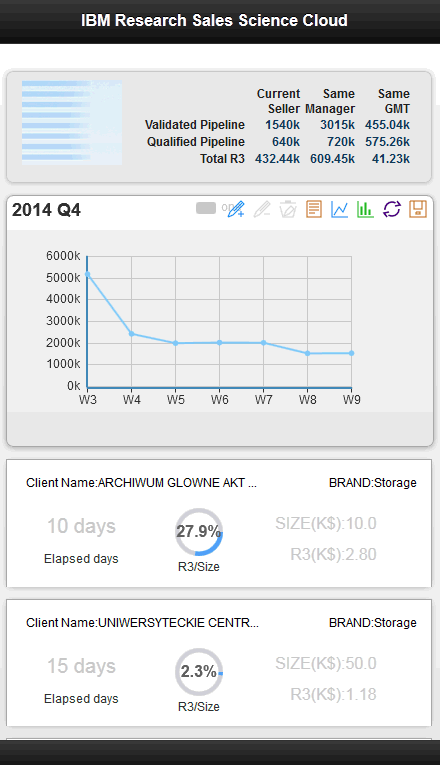}
\includegraphics[width=0.147\textwidth]{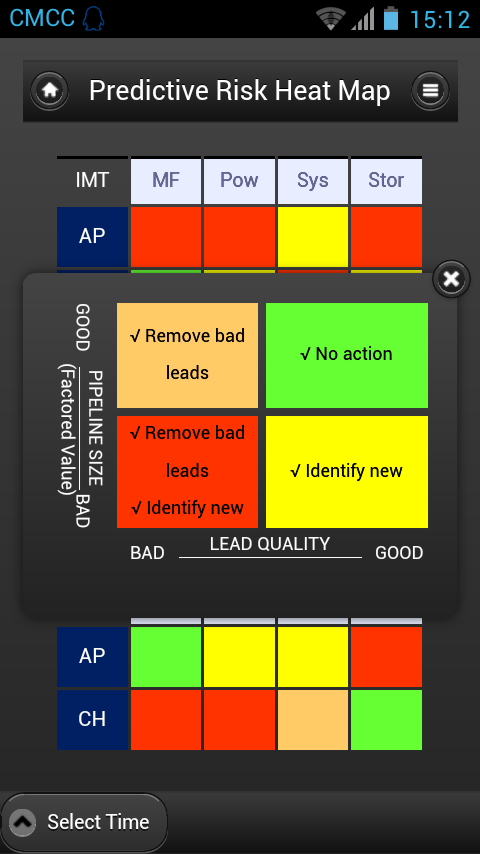}
\includegraphics[width=0.147\textwidth]{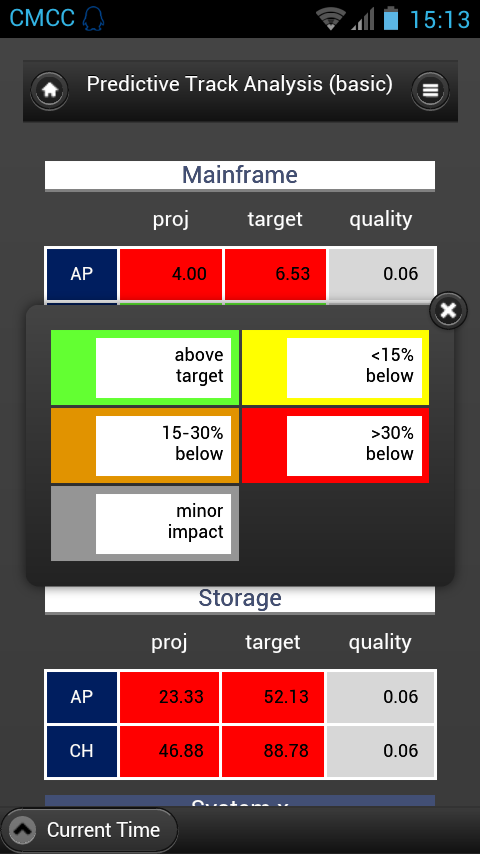}
\caption{Mobile application `Sales Assistant' for pipeline monitoring. Sellers can directly access the first-hand data together with predictive analytics results via the mobile application.}
\label{fig:app_snapshot}
\vspace{-5pt}
\end{figure}

\textbf{Mobile application} Moreover, we come up with a mobile version of the web portal. The aim is to provide the sales team, especially field sellers, who are mostly working in a client-facing environment, and use their mobile phone more frequently than laptop. In particular, the hope is not only the sellers would leverages our tools more conveniently, but also they are more willing to input additional information such as client-facing logs into the data system. This is a rich information worth integrating with the current pipeline data to improve the prediction accuracy. Several exemplary demo pages are shown in Figure \ref{fig:app_snapshot}.

The evaluation is designed as follows: first we collect the historical data for 13Q1, 13Q2, 13Q3, 13Q4, 14Q1, 14Q2 with known outcome in the corresponding quarter end.
Then we build two models for testing dataset of 14Q1 and 14Q2, respectively. For 14Q1, the training dataset includes the samples from 13Q1 and 13Q4, where 13Q1 accounts for the seasonality, and 13Q4 for the recency.
For the same reason, for 14Q2, the training set is composed from the combination of 13Q2 and 14Q1. Following this protocol, we use the gain score as the standard metric for comparing the performance both across geometries and weeks within the quarter.
In addition, sales subjective rating is also evaluated as a baseline. We consider the ``pure'' service, which denotes the associated hardware or/and software where the service is hosted also comes from the same vendor i.e. the referred company. Evaluation results are shown in Table \ref{tab:pure_14Q1}, \ref{tab:pure_14Q2}, where the average score of the model/seller follows the scores over thirteen weeks.
\begin{table}[tbp]
  \centering
  \caption{Gain score on 2014Q1 pure-service leads.}
    \vspace{-5pt}
    \resizebox{0.47\textwidth}{!}{
    \begin{tabular}{ccccccccc}
    \toprule
   Week&GCG&Japan&AP&LA&CEE&NA&MEA&EU\\
    \midrule
      1&0.706&0.529&0.782&0.721&0.671&0.724&0.806&0.674\\
2&0.728&0.574&0.81&0.71&0.671&0.735&0.783&0.679\\
3&0.717&0.577&0.823&0.712&0.7&0.576&0.792&0.69\\
4&0.666&0.624&0.837&0.702&0.674&0.531&0.798&0.702\\
5&0.587&0.671&0.849&0.707&0.666&0.563&0.800&0.71\\
6&0.605&0.87&0.846&0.724&0.659&0.564&0.769&0.727\\
7&0.606&0.866&0.848&0.732&0.659&0.592&0.785&0.745\\
8&0.672&0.866&0.84&0.718&0.604&0.611&0.785&0.74\\
9&0.691&0.800&0.839&0.69&0.62&0.619&0.796&0.741\\
10&0.715&0.736&0.822&0.708&0.623&0.682&0.825&0.754\\
11&0.696&0.84&0.835&0.701&0.688&0.663&0.824&0.769\\
12&0.75&0.827&0.822&0.703&0.752&0.702&0.827&0.795\\
13&0.778&0.963&0.845&0.764&0.799&0.738&0.846&0.81\\
model&\textbf{0.686} &\textbf{0.749} &\textbf{0.831} &\textbf{0.715} &\textbf{0.676} &\textbf{0.638} &\textbf{0.803} &\textbf{0.734}\\
    \midrule
    1&0.523&0.706&0.524&0.666&0.686&0.738&0.726&0.616\\
2&0.509&0.709&0.57&0.591&0.681&0.712&0.68&0.609\\
3&0.507&0.708&0.57&0.579&0.665&0.643&0.676&0.621\\
4&0.483&0.688&0.566&0.619&0.672&0.645&0.68&0.624\\
5&0.482&0.671&0.58&0.576&0.667&0.675&0.675&0.615\\
6&0.481&0.682&0.579&0.597&0.687&0.706&0.7&0.63\\
7&0.478&0.711&0.569&0.578&0.647&0.718&0.702&0.633\\
8&0.482&0.708&0.557&0.556&0.675&0.733&0.705&0.625\\
9&0.455&0.754&0.561&0.56&0.714&0.747&0.711&0.634\\
10&0.452&0.79&0.574&0.553&0.726&0.743&0.709&0.633\\
11&0.47&0.709&0.612&0.568&0.707&0.744&0.677&0.642\\
12&0.535&0.789&0.578&0.561&0.688&0.738&0.669&0.635\\
13&0.645&0.944&0.556&0.462&0.708&0.721&0.68&0.645\\
seller&\textbf{0.500} &\textbf{0.736} &\textbf{0.569} &\textbf{0.574} &\textbf{0.686} &\textbf{0.713} &\textbf{0.692} &\textbf{0.628} \\
    \bottomrule
    \end{tabular}}%
  \label{tab:pure_14Q1}%
\end{table}
\begin{table}[tbp]
  \centering
  \caption{Gain score on 2014Q2 pure-service leads.}
    \vspace{-5pt}
    \resizebox{0.47\textwidth}{!}{
    \begin{tabular}{ccccccccc}
    \toprule
   Week&GCG&Japan&AP&LA&CEE&NA&MEA&EU\\
    \midrule
      1&0.614&0.718&0.764&0.646&0.538&0.722&0.618&0.665\\
2&0.656&0.75&0.77&0.677&0.517&0.74&0.657&0.677\\
3&0.717&0.622&0.778&0.764&0.539&0.742&0.68&0.679\\
4&0.725&0.689&0.778&0.779&0.572&0.734&0.668&0.708\\
5&0.762&0.75&0.764&0.778&0.554&0.758&0.692&0.718\\
6&0.774&0.771&0.768&0.785&0.593&0.753&0.705&0.723\\
7&0.79&0.725&0.753&0.791&0.589&0.784&0.718&0.733\\
8&0.804&0.75&0.753&0.795&0.638&0.791&0.739&0.747\\
9&0.83&0.833&0.772&0.804&0.648&0.804&0.771&0.733\\
10&0.838&0.878&0.756&0.762&0.656&0.802&0.784&0.734\\
11&0.842&0.9&0.759&0.741&0.681&0.818&0.841&0.759\\
12&0.844&0.871&0.775&0.751&0.737&0.815&0.83&0.778\\
13&0.84&0.824&0.783&0.737&0.774&0.814&0.827&0.79\\
model&\textbf{0.772} &\textbf{0.775} &\textbf{0.767} &\textbf{0.755} &\textbf{0.618} &\textbf{0.775} &\textbf{0.733} &\textbf{0.726} \\
    \midrule
1&0.561&0.405&0.583&0.597&0.707&0.65&0.563&0.594\\
2&0.566&0.312&0.586&0.612&0.699&0.664&0.592&0.603\\
3&0.565&0.439&0.582&0.652&0.684&0.655&0.589&0.658\\
4&0.554&0.439&0.574&0.657&0.684&0.637&0.604&0.69\\
5&0.566&0.305&0.576&0.644&0.671&0.648&0.629&0.681\\
6&0.57&0.406&0.575&0.671&0.663&0.661&0.604&0.68\\
7&0.573&0.406&0.585&0.702&0.681&0.672&0.599&0.692\\
8&0.574&0.406&0.587&0.723&0.686&0.688&0.61&0.694\\
9&0.584&0.412&0.589&0.732&0.679&0.696&0.597&0.673\\
10&0.579&0.4&0.601&0.714&0.687&0.703&0.61&0.664\\
11&0.587&0.325&0.625&0.667&0.681&0.706&0.64&0.657\\
12&0.581&0.329&0.601&0.686&0.675&0.701&0.673&0.647\\
13&0.529&0.441&0.609&0.674&0.684&0.685&0.666&0.635\\
seller&\textbf{0.568} &\textbf{0.387} &\textbf{0.590} &\textbf{0.672} &\textbf{0.683} &\textbf{0.674} &\textbf{0.614} &\textbf{0.659} \\
    \bottomrule
    \end{tabular}}%
  \label{tab:pure_14Q2}%
\vspace{-5pt}
\end{table}

\section{Conclusion and Outlook}
In this paper, we have described a predictive solution for data-driven sales opportunity win propensity prediction. The scoring models are built by using training examples drawn from historical transactions and explanatory features extracted from transactional data. Specific aspects in the sales pipeline are discussed and studied. For the ongoing work, we are modeling and forecasting of the customers' purchasing event sequence, such as self/mutual-exciting point process models \cite{YanIJCAI13,YanAAAI15}. Also, the sales resource optimization problem can be further studied \cite{WangEJOR13}, especially by applying several state-of-the-art graph matching methods \cite{TianECCV12,YanICCV13,YanECCV14,YanAAAI15P,YanTIP15}.

\textbf{Acknowledgment} The authors would like to thank Dr. Yasuo Amemiya with IBM T.J. Watson Research Center who shares the data for the experiments. This paper is partially supported by NSFC-61129001, 61025005, 61221001.

\end{document}